\documentclass[12pt]{article}% добавляется twoside для двусторонней печати

\newlength{\uppermar}
\newlength{\lowermar}
\newlength{\leftmar}
\newlength{\rightmar}

\usepackage[centertags]{amsmath}
\usepackage{amsfonts}
\usepackage{amssymb}
\usepackage{amsthm}
\usepackage{calc}

\begin{document}

\title{\textbf{Deformation
quantization \\ of linear dissipative systems}}

\author
{ V.G. Kupriyanov, S.L. Lyakhovich and A.A. Sharapov\\
\textit{Physics Department, Tomsk State University, Tomsk, 634050,
Russia} }

\maketitle

\renewcommand{\thefootnote}{\fnsymbol{footnote}}
\footnotetext{\small The work benefited from the following
research grants:  Russian Ministry of Science and Education grant
130337 and the grant for Support of Russian Scientific Schools
1743.2003.2. VGK and AAS appreciate the financial support from
Dynasty Foundation and International Center for Fundamental
Physics in Moscow. SLL is partially supported by RFBR grant
05-01-00996; AAS is partially supported by RFBR grant
03-02-17657.}
\renewcommand{\thefootnote}{\arabic{footnote}}
\setcounter{footnote}{0}

%%%%%%%%%%%%%%%%%%%%%%%%%%%%%%%%%%%%%%%%%%%%%%%%%%%%%%%%%
\begin{abstract}

    A simple pseudo-Hamiltonian formulation is proposed for the linear inhomogeneous systems of
    ODEs. In contrast to the usual Hamiltonian mechanics, our
    approach is based on the use of a non-stationary Poisson
    bracket, i.e. the corresponding Poisson tensor is allowed to explicitly depend on time.
    Starting from this pseudo-Hamiltonian formulation we develop
    a consistent deformation quantization procedure involving a non-stationary star-product $*_t$
    and an ``extended'' operator of time derivative
    $D_t=\partial_t+\cdots$, differentiating the $\ast_t$-product.
     As in the usual case, the $\ast_t$-algebra of
     physical observables is shown to admit an essentially unique
     (time dependent) trace functional $\mathrm{Tr}_t$.
    Using these ingredients we construct a complete and fully consistent
    quantum-mechanical description for any linear dynamical system with or without dissipation.
   The general quantization method is exemplified by the models of damped oscillator
   and radiating point charge.

\end{abstract}

%%%%%%%%%%%%%%%%%%%%%%%%%%%
\section{Introduction}

The problem of quantum-mechanical description of dissipative
systems remains for decades a recurrently discussed physical topic
with a number of important applications (see
\cite{Bateman}-\cite{Mensky} and references therein). It has also
some theoretical importance as a touchstone for testing various
quantization methods. Our interest to the problem is inspired by
recent developments in deformation quantization \cite{DS}. In this
paper we consider the problem of deformation quantization for
dissipative systems, whose classical dynamics is specified by
linear inhomogeneous systems of ODEs.

In spite of a large number of papers devoted to the
quantum-mechanical treatment of various dissipative  systems there
is no commonly accepted definition of the dissipation phenomenon
itself. It seems that the most characteristic property of the
dissipation, shared by all the systems known as dissipative, is
the presence of \textit{attractors} \cite{sb}. Recall that an
attractor $A\subset M$ is a compact, measure-zero subset in the
phase space of the system  $M$,  possessing the property of being
limiting set for  any  trajectory passing through a sufficiently
small neighborhood of $A$. In a simple situation the attractor is
just a stable fixed point (sink) or a closed curve (limit cycle);
in the general case, however, much more complicated dissipative
structures may occur (e.g. strange attractors).

\ From the Liouville theorem about the conservation of phase-space
volume \cite{Arnold} it follows immediately that no smooth
compatible symplectic structure can exist in the vicinity of an
attractor: As the measure of attractor is zero, the Liouville
volume form, given by the Pfaffian  of the symplectic 2-form,
becomes necessarily infinite at the points of the attractor. The
attractors are thus obstructions for constructing the conventional
Hamiltonian description of the dissipative systems and this is
also the reason why the terms ``dissipative'' and
``non-Hamiltonian'' are often used as synonyms.

In principle, the requirement of smoothness  of the symplectic
structure is not so crucial for deformation quantization  as only
the Poisson bracket and a trace density are actually used.  The
problem, however, is that even for a linear dissipative system
with one-point attractor (e.g. focus) the corresponding Poisson
bracket appears to be highly nonlinear, making impossible any
practical calculations \cite{kup} (see also \cite{BFFLS} for
similar discussion of Kepler's problem).

To get round the ``no-go theorem'' above and obtain a practicable
quantization scheme  we allow the symplectic structure to depend
on time explicitly. The idea is as follows: Since the phase-space
trajectories reach the attractor $A$ only asymptotically (as
$t\rightarrow \infty$) one can try to construct a time dependent
symplectic structure $\Omega(t)$ which would be a smooth function
of time and such that $\lim_{t\rightarrow\infty}\Omega(t)=\infty$.
We show that such a non-stationary symplectic structure does exist
for any linear dynamical system and can be explicitly constructed
by the fundamental matrix of the linear system. Moreover, for
one-dimensional dynamical systems, described by a linear
second-order equation, one can always find a time-independent
$\Omega$, whereas for the general multi-dimensional system such
choice is impossible. In this respect the one-dimensional
dissipative systems, like the damped linear oscillator, are not
indicative examples.

The unavoidable time dependence of multi-dimensional symplectic
structure has further consequences for nearly Hamiltonian systems,
i.e. the systems whose equations of motion have the form of a
small perturbation over the Hamiltonian ones. No matter how small
the value of perturbation is, the  perturbed equations may no
longer be Hamiltonian, that indicates presence of genuine
dissipation. It is still possible, however, to construct the
first-order action functional if one admits a non-stationary
symplectic structure; in so doing, the explicit time dependence of
$\Omega$ can violate any phase-space polarization
\cite{Woodhouse}(e.g., this may admit no separation of the
variables onto position coordinates and conjugate momenta) making
impossible the passage to the second-order Lagrangian formalism.
The absence of the second-order Lagrangian description for some
classical systems with two or more degrees of freedom was
recognized long ago \cite{Duglas}, \cite{Dodonov},
\cite{Henneaux}. For a nearly Hamiltonian system it is reasonable
to require the Lagrangian, if any, to pass into the non-perturbed
one as the parameter of dissipation (e.g. friction constant)
vanishes. The last requirement imposes very strong restriction on
physically reasonable  Lagrangians. For example, in \cite{kup} we
have described all the second-order Lagrangians for the reduced
Lorentz-Dirac equation in the case of homogeneous magnetic field,
but none of them reproduces the free particle motion as the charge
of the particle tends to zero. The last fact shows clearly the
advantage of the first-order formalism over the second-order one.
For the general discussion on peculiarities of the inverse problem
of variational calculus in the first-order formalism we refer the
reader  to \cite{Hojman}.

The absence of a phase-space polarization favors also the use of
deformation quantization over the canonical quantization
procedure. Being the function of time, but not of the phase-space
coordinates, the corresponding Poisson bracket can easily be
quantized by the usual Weyl-Moyal formula giving rise to a
\textit{non-stationary} star-product $*_t$. It is the point where
our approach deviates from the conventional scheme of deformation
quantization \cite{BFFLS}, \cite{Fedosov}. Notice that in the
non-stationary case the usual time derivative does not
differentiate the $\ast_t$-product of quantum observables; instead
one can define an extended time derivative $D_t=\partial_t+\cdots$
that would be compatible with the $\ast_t$-product in the sense of
the Leibnitz rule. As in the usual case, the $\ast_t$-algebra of
quantum observables is shown to admit an essentially unique
(non-stationary) trace functional $\mathrm{Tr_t}$. Using these
ingredients, we define the quantum Liouville equation governing
the evolution of quantum-mechanical states, as well as the rule
for computing the expectation values of physical observables. As a
result we get a complete quantum-mechanical description for any
linear dynamical system.  Moreover, the physical content of the
theory is shown to be independent of any ambiguities concerning
the choice of the non-stationary symplectic structure $\Omega(t)$
as the quantum Liouville equation coincides precisely with the
classical one (see Proposition 1). Notice that a lot of objections
have been published for many years against possibility to
consistently quantize such systems in the Hamiltonian framework,
even in principle, see e.g. \cite{Britin}-\cite{Mensky}.

By way of illustration we consider the quantization problem for
the damped linear  oscillator and for the radiating point charge
moving in a homogeneous magnetic field. In both cases the quantum
dynamics seems very reasonable. In particular, the time evolution
of mean energy, defined in terms of the corresponding unperturbed
system, is shown to coincide with evolution of the classical
energy.

\section{Pseudo-Hamiltonian formulation of linear dynamical systems}

We start with an inhomogeneous linear system of ODEs
\begin{equation}\label{1}
    \dot x^i=A^i_j(t)x^j+J^i(t)
\end{equation}
defined on a linear phase space with coordinates $x^i$. Hereafter
the overdot  stands for the derivative with respect to time $t$.
When $A$ and $J$ are independent of time, one speaks of an
autonomous system of ODEs. In this paper we are interested in
linear dynamical systems which are a (small) perturbation of  a
Hamiltonian one. Although the perturbed system may no longer  be a
Hamiltonian one\footnote{In this case, the perturbation can't be
induced by a perturbation of the Hamiltonian. }, the number of
degrees of freedom remains the same. In particular, we will always
assume the phase space of the system (\ref{1}) to be even
dimensional, i.e. $i=1,...,2n$.

Our first observation is that any such system can be derived from
the variation principle for a quadratic action functional if the
explicit time dependence is admitted in the integrand. Consider
the following ansatz:
\begin{equation}\label{2}
    S[x]=\frac12\int dt(x^i\Omega_{ij}(t)\dot
    x^j-x^iB_{ij}(t)x^j-2C_i(t)x^i)\,,
\end{equation}
where
\begin{equation}\label{3}
    \Omega_{ij}=-\Omega_{ji}\,,\;\;\;\;\;B_{ij}=B_{ji}\,,\;\;\;\;\;\det(\Omega_{ij})\neq
    0\,.
\end{equation}
Structurally, the functional $S$ is similar to the first-order
action associated with the Hamiltonian
\begin{equation}\label{Ham}
H=\frac12 x^iB_{ij}(t)x^j-C_i(t)x^i\,,
\end{equation}
but unlike the usual Hamiltonian formalism we allow the symplectic
form $\Omega$ to depend on time. This assumption appears to be
crucial for the description of multi-dimensional dissipative
systems as will be seen below.

Varying this action functional, we come to the following equations
\footnote{Here we use the matrix notation.}:
\begin{equation}\label{4}
\frac{\delta S}{\delta x^i}=0\,\;\;\;\Leftrightarrow\, \;\;\;
\dot{x} =\Omega^{-1}\left(
B-\frac12\dot{\Omega}\right)x+\Omega^{-1}C\,.
\end{equation}
In order for these equations to be equivalent to the original ones
(\ref{1}) we must set
\begin{equation}\label{6}
A=\Omega^{-1}\left(B-\frac12\dot{\Omega}\right)\,,\quad
J=\Omega^{-1}C\,,
\end{equation}
or, what is the same,
\begin{equation}\label{7}
\frac12\dot{\Omega}=B-\Omega A\,,\qquad C=\Omega J\,.
\end{equation}
Decomposing the first matrix equation onto symmetric and
anti-symmetric parts, we finally get
\begin{equation}\label{8}
   \dot{\Omega}=-(\Omega A+A^t\Omega) \,,\qquad  B=\frac12\biggl(\Omega A-A^t\Omega
   \biggr)\,,\qquad C=\Omega J\,,
\end{equation}
$A^t$ being the transpose of the matrix $A$. Only the first
relation is nontrivial (it is a linear ODE on $\Omega$), while the
other two are just definitions of the matrices $B$ and $C$.

Recall that the square matrix $\Gamma(t)$ is called the
fundamental solution to the system (\ref{1}) if
\begin{equation}\label{}
\dot \Gamma=A\Gamma \,, \qquad \Gamma(0)=1\,.
\end{equation}
The columns of this matrix constitute the basis in the linear
space of solutions to Eqs.(\ref{1}). Given the matrix $\Gamma$,
the general solution to the first equation (\ref{8}) can be
written as
\begin{equation}
\Omega=\Lambda^t\Omega_0\Lambda\,,
\end{equation}
where $\Lambda = \Gamma^{-1}$, and $\Omega_0=-\Omega_0^t$ is a
constant non-degenerate matrix. The matrix $\Omega_0$ encodes all
the ambiguity in the definition of quadratic action functional
(\ref{2}) for the given system of ODEs (\ref{1}).

In the autonomous case, the system (\ref{1}) is explicitly
integrable in elementary functions and hence the action (\ref{2})
can be written in a closed form. By way of illustration let us
consider two physical examples: the damped linear oscillator, and
the non-relativistic motion of a point charge with a due regard
for the radiation back reaction.

\vspace{5mm}\noindent {\textbf{Example 1.}} The equation
describing the linear oscillator with friction reads
\begin{equation}\label{dos}
    \ddot x+2\alpha x +\omega^2x=0\,.
\end{equation}
Here $\omega$ is the frequency and $\alpha\geq 0$ is the
coefficient of friction. Introducing  the auxiliary variable
$$p=\frac{\dot x+{\alpha} x}{\sqrt{1-\alpha ^2/\omega ^2}}\;,$$
one can replace (\ref{dos}) with the following pair of first-order
equations:
\begin{equation}
\dot x=p\sqrt{1-\alpha ^2/\omega ^2}-\alpha x\,,\qquad \dot
p=-\omega^2x\sqrt{1-\alpha ^2/\omega ^2}-{\alpha} p\,. \label{31}
\end{equation}
According to (\ref{2}) and (\ref{8}), the action functional to
this system is given by
\begin{equation}
 S[x,p]=c\int dt\left(p\dot x-\frac12\sqrt{1-\alpha ^2/\omega ^2 }(p^2+\omega^2x^2)+\alpha px\right)e^{2\alpha
 t}\,,
\label{32}
\end{equation}
$c$ being an arbitrary constant. In the regime of aperiodic
damping ($\alpha > \omega$) the action functional becomes complex.

In this simple case it is also possible to construct a first-order
action functional involving the canonical  symplectic structure.
For example, varying the action
\begin{equation}\label{32a}
 S[x,p]=\int dt\left(p\dot x-\frac12(e^{-\alpha t} p^2+\omega^2 e^{\alpha
 t}x^2)\right)\,,
\end{equation}
one gets the Hamiltonian equations
\begin{equation}\label{31a}
\displaystyle \dot x=e^{-\alpha t} p\,,\qquad \dot p= -\omega^2
e^{\alpha t} x\,,
\end{equation}
which are obviously equivalent to Eq.(\ref{dos}). Notice that,
contrary to the system (\ref{31}), the stationary point $x=p=0$ of
(\ref{31a}) is not an attractor.

It is significant that both
 (\ref{32}) and (\ref{32a}) come to the standard Hamiltonian action for the
harmonic oscillator when $\alpha\rightarrow 0$. Since the action
functional carries all the information about classical and quantum
dynamics, the last fact makes possible a consistent interpretation
of friction as a small perturbation over the given Hamiltonian
system, rather than something leading to a completely different
physical system.

\vspace{5mm}\noindent {\textbf{Example 2.}} The effective dynamics
of a non-relativistic charged particle is governed by the Lorentz
equation \cite{LL}
\begin{equation}\label{f4}
m\ddot{\bf\rm x}=e{\bf E}+\frac ec [\dot{\bf\rm x},{\bf H}]+
\frac{2e^2}{3c^3}\stackrel{...}{\rm\bf x}\,.
\end{equation}
Here ${\bf\rm x}(t)\in \mathbb{R}^3$ is a trajectory of the
particle,  $\mathbf{E}$ and $\mathbf{H}$ are 3-vectors of electric
and magnetic fields, and  the constants $c$ and $e$  denote the
light velocity and electric charge of the particle. As is seen,
the Lorentz equation involves third time derivative of the
trajectory. It is the term which describes the back reaction of
the radiation emitted by the accelerating charge. Since the order
of the equation is greater than two, it cannot be assigned with a
straightforward mechanical interpretation: in the realm of
Newtonian mechanics, a trajectory of a scalar particle is uniquely
specified by initial position and velocity. Besides, together with
physically meaningful  solutions  Eq.(\ref{f4}) allows a set of
nonphysical ones \cite{LL}. It turns out that both mentioned
problems can be resolved by means of the \textit{reduction of
order procedure} (see e.g. \cite{kup}, \cite{LL}, \cite{KSh}).
Namely, Eq. (\ref{f4}) is replaced by a second-order equation
$\ddot{\mathbf{x}}=\mathbf{g}(\mathbf{x}, \dot{\mathbf{x}}, e)$
such that all the solutions to the latter would solve the former.
The last requirement leads to  a partial differential equation on
the function $\mathbf{g}(\mathbf{x},\mathbf{v}, e)$ having a
unique solution with $\mathbf{g}(\mathbf{x},\mathbf{v},0)=0$.

Consider for example  a non-relativistic particle moving in a
homogeneous magnetic field, i.e. $\mathbf{E}=0$, $\mathbf{H}=(0,0,
H)$ and $H=const$. The reduced Lorentz equation has the form
\cite{kup}
\begin{equation}
\ddot x=A\dot x-B\dot y\,,\qquad \ddot y=B\dot x+A\dot y\,,\qquad
\nonumber \ddot z=0 \,, \label{f19}
\end{equation}
where $\mathbf{x}=(x,y,z)$ and
\begin{equation}
A=\frac{6-\sqrt{6}\sqrt{3+\sqrt{9+64e^6H^2}}}{8e^2}\approx
-\frac{2}{3}e^4H^2\,,\;\;\;\;\;\;
B=\frac{eH\sqrt{6}}{\sqrt{3+\sqrt{9+64e^6H^2}}}\approx eH\,.
\end{equation}
Here we have  set $m=c=1$. Since the evolution along $z$
represents the free motion and decouples from the dynamics in the
$xy$-plane, we restrict our consideration to the first two
equations.

Setting  formally $A=0$, we arrive at the usual Lorentz equations
describing the motion of a charge in response to the ``effective''
magnetic field  ${\bf B}=(0,0,B/e)$. In this case, the
trajectories are concentric circles. For $A\neq 0$ the particle
spirals at the origin of $xy$-plane. So, it is natural to regard
$A$ as the coefficient of friction.

In order to construct an action functional for the second-order
Eqs.(\ref{f19}), we  replace them with an equivalent system of
first-order ones. Let us choose the auxiliary variables as
\begin{equation}
p=\dot x +\frac{B}{2} y ,\indent q=\dot y-\frac{B}{2} x\,.
\end{equation}
Then
\begin{equation}
\begin{array}{ll}
\displaystyle \dot x=p-\frac{B}{2} y\,,&\displaystyle \dot
y=q+\frac{B}{2} x\,,\\  [5mm] \displaystyle \dot p=-\frac{B}{2}
q-\frac{B^2}{4} x+ A\left(p-\frac{B}{2} y\right)\,,\qquad&
\displaystyle \dot q=\frac{B}{2} p-\frac{B^2}{4}
y+A\left(q+\frac{B}{2} x \right)\,. \label{f39}
\end{array}
\end{equation}
Applying the general formulas (\ref{2}) and (\ref{8}), we arrive at the following
expression for the first-order action functional:
\begin{equation}\label{47}
 S[x,y,q,p]=\frac{1}{4(A^2+B^2)}\int dte^{-At}[2a(t)(p\dot
x-x\dot p +q\dot y-y\dot q)
\end{equation}
$$
+2b(t)(q\dot x-x\dot q+y\dot p-p\dot y)+ 2c(t)(p\dot q-q\dot
p)+2d(t)(x\dot y -y\dot x)
$$

$$+e(t)(p^2+q^2)+ f(t)(x^2+y^2)+g(t)(px+qy) +j(t)(qx-py)]\,,
$$
where
$$
a(t)=A^2\cos(Bt)+\frac12 B^2 (e^{-At}+e^{At})\;,\;\;\;\;
b(t)=A^2\sin(Bt)-ABe^{At}+AB \,\cos(Bt)\;,
$$ $$
c(t)=e^{-At}B+2A\,\sin(Bt)\;,\;\;\;
d(t)=\frac14B^3(e^{-At}-e^{At})-\frac12B^2A\,\sin(Bt)+A^2B(\cos(Bt)-e^{At})\;,
$$ $$
e(t)=e^{-At}B^2+A^2\cos(Bt)+A\,\sin(Bt)B\;,
$$ $$
f(t)=\frac14B\left(B^3e^{-At}+BA^2\cos(Bt)
-A\,\sin(Bt)[B^2+2A^2]\right)\;,$$ $$
g(t)=-A\,\cos(Bt)B^2-A^3\cos(Bt)\;,\;\;\;
j(t)=-A^3\sin(Bt)+e^{-At}B^3+A^2B\,\cos(Bt)\;.
$$
In the limit of zero friction $A\rightarrow 0$, this  action tends
to the usual action for the charged particle in a homogeneous
magnetic field $B$.

Notice that the symplectic structure entering the Lagrangian
(\ref{47}) does not possess  $xy$-polarization (i.e. the 2-form
$\Omega$ does not vanish upon restriction on the 2-plane $x=const,
y=const$) when $A\neq 0$. This makes impossible the algebraic
elimination of $p$ and $q$ from the action (\ref{11}) and thus
obtaining a second-order action in terms of $x$ and $y$. The
latter fact agrees well with the general statement of
Ref.\cite{kup} about nonexistence of a second-order action
functional for Eq. (\ref{f19}), which would pass to the standard
action functional for a free particle when $e\rightarrow 0$.

It can also be shown that unlike the previous example, the system
(\ref{f19}) does not admit a first-order action involving a
stationary symplectic structure and having the standard free
limit. The unavoidable time dependence of the symplectic structure
may be thus viewed as a specific feature of multi-dimensional
dissipative systems.

\vspace{5mm}

Using the Hamiltonian (\ref{Ham}) and the symplectic 2-form $\Omega$ it is
possible to rewrite  Eqs.(\ref{1}) in a pseudo-Hamiltonian form.
To this end we introduce the following nonstationary Poisson
brackets:
\begin{equation}\label{9}
    \{F,G\}_t=\Pi^{ij}(t)\frac{\partial F}{\partial
    x^i}\frac{\partial G}{\partial
    x^j}\,,\;\;\;\;\;\Pi=\Omega^{-1}\,,
\end{equation}
where $F$ and $G$ are functions of the phase-space coordinates
$x^i$ and the time $t$. Clearly, these brackets satisfy all the
properties of the Poisson brackets: bi-linearity, skew-symmetry,
the Leibnitz and Jacobi identities. In accordance with (\ref{4})
the evolution of an arbitrary physical observable $F(t,x)$ is
described by the following equation:
\begin{equation}\label{10}
\frac{d F}{dt}=D_t F + \{F,H\}_t\,,
\end{equation}
where
\begin{equation}\label{h}
H=\frac12 x^iB_{ij}x^j+C_i(t)x^i
\end{equation}
and
\begin{equation}\label{11}
    D_tF \equiv\frac{\partial F}{\partial t}-\frac12
    x^i\dot{\Omega}_{ik}\Pi^{kj}\frac{\partial F}{\partial x^j}
   = \frac{\partial F}{\partial t}-\frac12
    x^i\dot{\Omega}_{ik}\{x^k,F\}_t
\end{equation}
is the ``extended'' partial derivative in $t$. When $\Omega$ is
constant, the last term in (\ref{11}) vanishes and we arrive at
the conventional Hamiltonian equations w.r.t. the canonical
Poisson brackets and  the Hamiltonian (\ref{h}). The main property
of the extended time derivative $D_t$ is that it differentiates
the non-stationary Poisson brackets (\ref{9}), i.e.
\begin{equation}\label{12}
    D_t\{F,G\}_t=\{D_tF,G\}_t+\{F,D_tG\}_t\,,
\end{equation}
for any $F(t,x)$ and  $G(t,x)$. Using this property one can deduce
the Poisson theorem:
\begin{equation}\label{13}
    \frac{d F}{dt}=0\,,\;\;\frac{d
    G}{dt}=0\,\;\;\;\;\Rightarrow\;\;\,\frac{d}{dt}\{F,G\}_t=0\,
\end{equation}
(the Poisson bracket of two conserved quantities is conserved).

In the full analogy with the conventional Hamiltonian mechanics, a
state of the system is described by a classical distribution
function $\rho_{cl}(x,t)$ subject to the normalization condition
\begin{equation}\label{norm}
    \int d\mu \rho_{cl}(x,t) = 1\,,
\end{equation}
$d\mu \equiv \sqrt {\det \Omega} \,d^{2n}x$ being the Liouville
measure associated to the non-stationary  symplectic form
$\Omega$. The time dependence of $\rho_{cl}(x,t)$ is determined by
the modified Liouville equation (cf. Eq. (\ref{10}))
\begin{equation}\label{Li}
D_t \rho_{cl} = \{H,\rho_{cl}\}_t\,.
\end{equation}
Note that the time evolution preserves  the normalization
condition (\ref{norm}). Indeed, using the obvious identity
\begin{equation}
\int d\mu\{F, G\}_t=0\,,
\end{equation}
where one of the functions $F(x)$ and $G(x)$ vanishes on infinity,
one can find
\begin{equation}\label{}
    \frac{d}{dt}\int d\mu \rho_{cl}(x,t)=
    \left(\frac{d\ln\Delta}{dt}-\dot\Omega_{ij}\Pi^{ij}\right)\int d\mu
    \rho_{cl}(x,t)=0\,,\qquad\Delta = \sqrt{\det\Omega}\,.
\end{equation}
The \textit{pure states} of the classical system correspond to the
$\delta$-distributions
\begin{equation}\label{}
    \rho_{cl}(x,t)=\Delta^{-1}(t)\delta^{2n}(x-x_0(t))\,,
\end{equation}
supported on the integral trajectories $x_0(t)$ of the  system
(\ref{1}).

As the final remark, let us note that the above pseudo-Hamiltonian
formalism is applicable  for arbitrary (not necessary quadratic)
Hamiltonians and can easily be derived/justified in the formalism
of constrained dynamics \cite{GT} applied to the first-order
action (\ref{2}).

\section{Deformation quantization of pseudo-Hamiltonian systems}

Example 2 of Sec.1 shows that the most characteristic feature of
multi-dimensional dissipative systems is the lack of a phase-space
polarization compatible with dynamics: Though at each time moment
$t_0$ one can split the phase-space variables $x^i=(q^a,p_b)$ on
``coordinates'' and ``momenta'' satisfying the canonical Poisson
bracket relations
\begin{equation}\label{}
    \{q^a,q^b\}_{t_0}=0\,,\quad \{p_a,p_b\}_{t_0}=0\,,\quad
    \{q^a,p_b\}_{t_0}=\delta_b^a\,,
\end{equation}
these relations may not hold true at the next time moment. This is
due to the explicit (and in many interesting cases unavoidable)
time dependence of the symplectic form $\Omega$ entering the
first-order action (\ref{2}). The absence of a natural
polarization favors the use of deformation quantization over the
canonical quantization procedure.

In the approach of deformation quantization the classical
observables (i.e. functions in the phase-space variables) are
identified with  symbols of operators \cite{BFFLS},
\cite{Fedosov}; in so doing, the pointwise  multiplication of
functions is replaced by an associative noncommutative
star-product. Using the non-stationary Poisson bracket (\ref{9}),
we define the time dependent $\ast_t$-product by the usual
Weyl-Moyal formula
\begin{equation}\label{14}
    (F\ast_t G)(x,t)\equiv \exp\left(\frac{i\hbar}{2}\Pi^{ij}(t)\frac{\partial^2}
    {\partial x^i\partial y^j}\right)F(t,x)G(t,y)|_{x=y}=F\cdot G+\frac{i\hbar}{2}\{F,G\}_t+O(\hbar^2)\,,
\end{equation}
It is clear that the time dependence of $\Pi$ does not affect the
associativity of the Weyl-Moyal star-product, so we have
\begin{equation}\label{}
(F\ast_t G)\ast_t H=F\ast_t(G\ast_t H)\,,\qquad\forall \,F,G,H\in
C^{\infty}(\mathbb{R}^{2n})\,.
\end{equation}

In order to define the notion of a quantum state we endow the
$\ast_t$-algebra with the following trace functional:
\begin{equation}\label{18}
    \mathrm{Tr}_t(F)= \frac{1}{(2\pi\hbar)^n}\int d\mu
    F(x)\,.
\end{equation}
The basic property of the trace (specifying it up to
multiplication on an arbitrary  function of $t$) is vanishing on
$\ast_t$-commutators, i.e.
\begin{equation}\label{19}
    \mathrm{Tr}_t([F,G]_t)=0\,,
\end{equation}
where $[F,G]_t\equiv F\ast_t G-G\ast_t F$ and at least one of the
functions $F$ and $G$ vanishes on the infinity together with all
its derivatives. Actually, the identity (\ref{19}) follows from
the stronger one: $\mathrm{Tr}_t(F\ast_t G)=\mathrm{Tr}_t(F\cdot
G)$.

A \textit{pure state}  of a quantum mechanical system is described
by a Wigner function $\rho (t,x)$ subject to the following
conditions:
\begin{equation}\label{20}
    \rho\ast_t\rho=\rho\,,\qquad\mathrm{Tr}_t(\rho)=1\,.
\end{equation}
The quantum counterpart of the classical Liouville equation,
governing the evalution of a quantum state,  reads
\begin{equation}\label{16}
     i\hbar D_t\rho +[\rho,H]_t=0\,,
\end{equation}
where the extended time derivative $D_t$, defined by Eq.
(\ref{11}), can also be written as
\begin{equation}\label{17}
D_t F\equiv \frac{\partial F}{\partial t}-\frac{1}{4i\hbar}\dot\Omega_{ij}
     (x^i\ast_t [x^j,\;F \; ]_t+[x^j,\;F \;]_t\ast_t x^i) \,.
\end{equation}
Note that Eq.(\ref{16}) makes sense for an arbitrary (not
necessary quadratic) Hamiltonians.  For a constant Poisson
bracket, (\ref{16}) reproduces the usual von Neumann's equation
for the symbol of statistical operator $\rho$. A simple direct
calculation shows that the operator $D_t$ differentiates the
$\ast_t$-product:
\begin{equation}\label{15}
    D_t(F \ast_t G)=(D_tF)\ast_t G+F\ast_t(D_t G)\,.
\end{equation}
As a consequence, every solution to Eq. (\ref{16}), satisfying the
idempotency condition (\ref{20}) at some initial time moment, will
satisfy this condition in all subsequent time moments. The
specific choice of the integration measure in the definition of
trace functional (\ref{18}) provides the conservation of the
normalization condition (\ref{20}). Indeed, using the evolution
equation for $\rho$ and the property (\ref{19}), one can find
\begin{equation}\label{}
    \frac{d}{dt}\mathrm{Tr}_t(\rho(t))=
    \left(\frac{d\ln\Delta}{dt}-\dot\Omega_{ij}\Pi^{ij}\right)\mathrm{Tr}_t(\rho(t))=0\,.
\end{equation}
This property specifies the form of the trace functional up to an
overall  constant.

In the Schr\"{o}dinger picture\footnote{ Of course, all the
constructions can be straightforwardly  reformulated  in the
Heisenberg picture.} the physical  observables  are considered to
be chosen once and for all at some initial time moment, say $
\;t=0$. Only the quantum states evolve according to Eq.
(\ref{16}). The expectation value of an observable $F(x)$ at the
time moment $t$ relative to a state $\rho(t)$ is given by
\begin{equation}\label{21}
    \langle F \rangle^t_\rho=\mathrm{Tr}_t(F(x)\ast_t\rho(t,x))\,,
\end{equation}
where the Wigner function $\rho$ obeys Eqs.(\ref{20}).

For quadratic Hamiltonians the equation (\ref{16}) coincides with
the modified Liouville equation (\ref{Li}).  In that case, the
evolution of an arbitrary quantum state $\rho$ is governed by the
linear first-order PDO
$$
\frac{\partial \rho}{\partial t}-\frac{1}{2}x^i\dot\Omega_{ij}\{ x^j,\rho\}
+ \{ \rho,H\} =0\;,
$$
for which the initial classical equations  (\ref{1}) play the role
of characteristics. So, we arrive at the following

\vspace{5mm}\noindent\textbf{Proposition 1.} {\it Let $x^i(t)=
\Gamma^i_j(t) x^j_0+v^i(t)$ be the general solution to the
classical equations of motion (\ref{1}) with
$x^i(t)|_{t=0}=x_0^i$, then the evolution of a quantum state
$\rho(t,x)$ is given by the expression
$$\rho(t,x)=\rho_0(\Lambda(t)[x- v(t)])\,,$$ where
$\Lambda(t)=\Gamma^{-1}(t)$ and the initial state
$\rho_0(x)=\rho(0,x)$ satisfies Eqs.(\ref{20}) at $t=0$.}

\vspace{5mm} Notice that the quantum evolution of linear systems
is completely determined by the classical one and does not depend
on any ambiguities concerning the choice of the quadratic action
functional (\ref{2}).

As we mentioned in Introduction the characteristic feature of
dissipation is the presence of attractors, i.e. invariant subsets
$A\in\mathbb{R}^{2n}$ to which all nearby trajectories converge.
In other words, any classical state supported at a sufficiently
small vicinity of $A$ evolves to a state supported at $A$ when
$t\rightarrow\infty$. As the next proposition shows the similar
phenomenon takes place at the quantum level as well (at least for
linear systems with one-point attractor).

\vspace{5mm}\noindent {\textbf{Proposition
2}.}\indent\textit{Given a system of differential equations $\dot
x = A(t)x$ having the origin $x=0$ as the global attractor, then
each Wigner's function $\rho(t,x)$ defines a $\delta$-shaped
sequence}
$$\phi_t(x)=\frac{\Delta(t)}{(2\pi\hbar)^n}\rho(t,x)\,,\qquad
\lim_{t\rightarrow \infty}\phi_t(x)=\delta(x)\,.$$

\vspace{5mm}\noindent \textbf{Proof}. Since $x=0$ is the global
attractor $\lim_{t\rightarrow \infty} x(t)=0$ for any solution
$x(t)$. On the other hand, $x(t)=\Gamma(t)x_0$, and hence
$\lim_{t\rightarrow \infty}\Gamma(t)=0$. According to Proposition
1, for any compactly supported function $F(x)\in
C^\infty_0(\mathbb{R}^{2n})$ we have
\begin{equation}
\int F(x)\phi_t(x)d^{2n}x=\frac{\Delta(t)}{(2\pi\hbar)^n}\int
F(x)\rho_0(\Lambda(t)x)d^{2n}x=\frac{1}{(2\pi\hbar)^n}\int
F(\Gamma(t)x)\rho_0(x)d^{2n}x\,.
\end{equation}
Taking limit, we finally get
$$
\langle F\rangle_\rho^\infty=\lim_{t\rightarrow
\infty}\frac{1}{(2\pi\hbar)^n}\int F(\Gamma(t)x)\rho_0(x)d^{2n}x=
F(0)\mathrm{Tr}_0(\rho_0)=F(0)\,.
$$
Thus, $\phi_t(x)$ is a $\delta$-shaped sequence.

\section{Examples}

By way of illustration let us consider the deformation
quantization of two interesting dissipative systems: the damped
linear oscillator and the radiating point charge in a homogeneous
magnetic field. The corresponding classical dynamics has been
discussed in Sec.2, including the time-dependent Poisson brackets.
In both cases, dissipation has the form of a perturbation over a
Hamiltonian system and we choose the basis quantum states as the
eigen-states for the energy (and angular momentum) of the
corresponding non-perturbed system. Using these states we then
consider the evolution for  the mean values of energy (and angular
momentum) in the presence of dissipation.

\subsection{Damped linear oscillator}
Let $\rho(x,p)$ be the Wigner function describing a pure state of
the harmonic oscillator ($\alpha =0$) with a definite value of
energy $E$. This amounts to saying that $\rho$ solves the
following eigen-value problem:
\begin{equation}
\begin{array}{l}
\displaystyle H\ast\rho\;=\;\rho\ast H=E\rho\;,\\
[3mm] \displaystyle\rho\ast\rho=\rho\,,\qquad\bar \rho=\rho\,,\qquad
\mathrm{Tr}(\rho)=1\;,
    \label{33}
    \end{array}
\end{equation}
where
$$
H=\frac{1}{2}\left(p^2+\omega^2 x^2 \right)\,,
$$
and the  $\ast$-product (\ref{14}) is defined at $t=0$ by the
canonical Poisson brackets $\{p,x\}=1$. Let us find $\rho$ from
(\ref{33}) following \cite{BFFLS}. The first (complex) equation in
(\ref{33}) is equivalent to the pair of the real ones
\begin{equation}
H\rho-\frac{\hbar^2}{4}\left(\omega^2 \frac{\partial^2
\rho}{\partial p^2}+\frac{\partial^2 \rho}{
\partial x^2}\right)=2E\rho\,,\qquad \{ H,\rho\}  =0\,. \label{34}
\end{equation}
The second equation implies that
\begin{equation}
\rho=\rho(H)\,.
 \label{36}
\end{equation}
Then the first equation yields
\begin{equation}
H\rho-\frac{\hbar^2}{4}({\rho}'' H+{ \rho}')=\frac{E}{\omega
}\rho\,. \label{37}
\end{equation}
Introducing new variables
$$
y=\frac{4}{\hbar\omega}H\;,\;\;\;\;\rho=e^{-y/2}f\,,
$$
one can bring Eq. (\ref{37}) to the form
\begin{equation}
{f}''y+(1-y) f'+\left(\frac{E}{\hbar\omega
}-\frac{1}{2}\right)f=0\,. \label{38}
\end{equation}
This equation is satisfied by Laguerre's polynomials $L_n(y)$,
provided
\begin{equation}\label{E}
E =\hbar\omega\left(n+\frac12\right)\,,\qquad n=0,1,2,...\,,
\end{equation}
and these are known to exhaust all its solutions resulting in
integrable Wigner's functions $\rho\in L^1(\mathbb{R}^2)$. Thus,
any eigen-value (\ref{E}) corresponds to the unique Wigner's
function
\begin{equation}
\rho_n(H)=C_n\;\mathrm{exp}\left(-\frac{2H}{\hbar \omega^2}\right)
L_n\left(\frac{4H}{\hbar \omega^2 } \right) \label{40}
\end{equation}
The constant $C_n=(-1)^n2/\omega $ is determined from the
normalization condition (\ref{33}). In Ref. \cite{BFFLS}, it was
shown that the sequence $\{\rho_n\}$  defines a complete set of
orthogonal projectors:
\begin{equation}
\rho_n\ast\rho_m=\delta_{mn}\rho_n\,,\qquad
\delta(x-x')\delta(p-p')=\sum_{n=0}^\infty\rho_n(x,p)\rho_n(x',p')\,.\label{41}
\end{equation}

Now let us return to the damped linear oscillator with action
(\ref{32}). The time evolution of the mean energy $H$ can  easily
be calculated using Proposition 1 and the formula (\ref{21}).
 We have
\begin{equation}
\langle H\rangle^t_{\rho_n}=\frac{\Delta(t)}{(2\pi\hbar)^n}\int
H(x)\rho_n(\Lambda(t)x)d^{2n}x=\frac{1}{(2\pi\hbar)^n}\int
H(\Gamma(t)x)\rho_n(x)d^{2n}x\,,
\end{equation}
where the initial state $\rho_n$ is one of the states (\ref{40})
 and $\Gamma(t)$ solves Eq. (\ref{31}) with $\Gamma(0)=1$. Since
\begin{equation}
H(\Gamma(t)\xi)=e^{-\alpha t}H(\xi)\,, \qquad \xi=(x,p)^t\;,
\end{equation}
we get
\begin{equation}
 \langle H\rangle^t_{\rho_n}=\frac{e^{-\alpha
t}}{2\pi\hbar}\int dpdx H\ast\rho_n=\frac{e^{-\alpha
t}E_n}{2\pi\hbar}\int dpdx \rho_n=e^{-\alpha t}E_n\,.
\end{equation}
As is seen the quantum evolution of the mean energy relative to
the eigen-states $\rho_n$ coincides exactly with the classical
one: in both cases the energy decreases  by exponential law.

\subsection{Particle in homogeneous magnetic field}
Consider the deformation quantization of the system (\ref{f19}). It is
convenient to identify the complete set of observables with the
energy $H$ and the angular momentum $L$ of the particle without
friction ($A=0$). Then the corresponding set of states with
definite values of $H$ and $L$ is determined by the equations
\begin{equation}
\begin{array}{rl}
\displaystyle H\ast\rho=\rho\ast H=E\rho\,, \quad&
 L\ast\rho=\rho\ast L=M\rho\,,\\[3mm]
\displaystyle\rho\ast\rho=\rho\,,\quad&\bar \rho=\rho\,,\quad \mathrm{Tr}(\rho)=1\;,
    \label{f40}
    \end{array}
\end{equation}
where $E,M\in \mathbb{R}$ and
\begin{equation}
H=\frac{1}{2}\left(p-\frac{B}{2}
y\right)^2+\frac{1}{2}\left(q+\frac{B}{2} x \right)^2\,,\qquad
L=py-qx\,.
\end{equation}
The $\ast$-product is defined by the formula (\ref{14}) at $t=0$
w.r.t. the canonical Poisson brackets: $\{p,x\}=\{q,y\}=1$ and the
other brackets vanish.

In order to solve (\ref{f40}) we introduce the following linear
canonical transformation  $(p,x;q,y)\rightarrow (P,X;Q,Y)$:
\begin{equation}
\begin{array}{ll}
\displaystyle P =p-\frac{B}{2}y\,,&\displaystyle
X=\frac1B\left(q+\frac{B}{2}x\right)\,,\\[5mm]
\displaystyle Q =\frac1B\left(q-\frac{B}{2}x\right)\,,&
\displaystyle Y =p+\frac{B}{2}y\,.
    \label{f42}
    \end{array}
\end{equation}
together with the functions
\begin{equation}\label{}
    H_1=P^2+B^2X^2\,,\qquad H_2=Q^2+B^2Y^2\,.
\end{equation}
It easy to see that
\begin{equation}\label{}
    H=H_1\,,\qquad L=B^{-1}(H_2-H_1)\,.
\end{equation}
Since $H_1$ and $H_2$ are nothing but the Hamiltonians of two
independent harmonic oscillators, we have reduced the eigen-value
problem (\ref{f40}) to the previous one (\ref{33}). The Wigner
functions solving Eqs.(\ref{f40}) are given by the (ordinary)
products
\begin{equation}
\rho_{E,M}=\rho_n(H_1)\rho_l(H_2)\,,
\end{equation}
where $\rho_m(H_{1,2})$ is defined by (\ref{40}) with $\omega=B$, and the
eigen-values $E$ and $M$ run trough the sets
\begin{equation}\label{}
E=\hbar B\left(n+\frac12\right)\,,\qquad M=\hbar (l-n) \,,\qquad
n,l=0,1,2,...
\end{equation}
As is seen the eigen-values of the angular momentum $L$ in a state
with definite energy $E$ are bounded from below by $-E/B+\hbar/2$.

Consider now the evolution of the mean values of $H$ and $L$.
Using the fundamental matrix of (\ref{f39}) and applying Proposition 1 we find
\begin{equation}
\begin{array}{c}
H(\xi,t)=H(\Gamma(t)\xi)=e^{2At}H(\xi)\,,\\[5mm]
L(\xi,t)=L(\Gamma(t)\xi)=L(\xi)+ \alpha(t)H(\xi) +\beta (t)K(\xi)
+\gamma (t)N(\xi)\,, \label{f53}
\end{array}
\end{equation}
where $\xi=(x,p,y,q)^{\mathrm{t}}$ and
\begin{equation}
\begin{array}{ll}
\displaystyle
\alpha (t)=\frac{2A^2e^{At}\mathrm{cos}(Bt)-A^2+B^2e^{2At}}{B(B^2+A^2)}\,,&
K=PQ+XY\,,
\\[5mm]
\displaystyle \beta
(t)=\frac{2A^2e^{At}\mathrm{cos}(Bt)+2Ae^{At}\mathrm{sin}(Bt)B
-2A^2}{B(B^2+A^2)}\,,& N=XQ-PY\,, \\[5mm]
\displaystyle \gamma
(t)=\frac{2A^2e^{At}\mathrm{cos}(Bt)B-2A^2e^{At}\mathrm{sin}(Bt)
-2AB}{B(B^2+A^2)}\,.
\end{array}
\end{equation}
This gives immediately
\begin{equation}
\langle H\rangle^t_{E,M}=\frac{e^{2At}}{(2\pi\hbar)^2}\int
d^{4}\xi
H(\xi)\ast\rho_{E,M}(\xi)=\frac{e^{2At}E}{(2\pi\hbar)^2}\int
d^{4}\xi \rho_{E,M}(\xi)=e^{2At}E\,.
\end{equation}
The mean energy of the particle decreases by exponential low just
as it behaves in the classical theory.

The invariance of the Wigner functions $\rho_{E,M}(\xi)$ under
reversions in $PX$- and $QY$-planes suggests that $\langle
K\rangle_{E,M}=\langle N\rangle_{E,M}=0$, and hence
\begin{equation}\label{62}
\langle L\rangle^t_{{E,M}}=\mathrm{Tr}_0(L(t)\ast\rho_{E,M})=M
-\alpha (t)E\,.
\end{equation}
The same arguments show that
\begin{equation}
\langle\, x\rangle^t_{E,M}=\langle\, p\rangle^t_{E,M}=\langle\,
y\rangle^t_{E,M}=\langle\, q\rangle^t_{E,M}=0\,.
\end{equation}
So, at each moment of time the measured values of coordinates and
momenta relative to the state $\rho_{E,M}$ are equal to zero.

Taking successive limits $t\rightarrow \infty$ and $A\rightarrow
0$ in (\ref{62}), we find that  the limiting  value of the angular
momentum for a small friction $| A|  \ll 1$ is given by $M-E/B$.
Since $\beta(t) \rightarrow 0$, $\gamma(t)\rightarrow 0$ as
$t\rightarrow \infty$, the same limiting value appears in the
classical theory as well.

\end{document}